\documentclass[aps,prl,twocolumn,showpacs,%
	amsmath,%
	reprint%
	,preprintnumbers,
	]{revtex4-1}

\usepackage{amsmath}
\usepackage{graphicx}

\usepackage{graphicx}
\usepackage{amssymb}
\usepackage{epstopdf}

\def\R{\mathrm R}
\def\C{\mathrm C}
\def\e{\mathrm e}
\def\S{\mathcal S}
\def\P{\mathcal P}

\begin{document}

\title{Threshold resonance and controlled filtering in quantum star graphs}

\author{Ond\v{r}ej Turek}
\author{Taksu Cheon}
\affiliation{Laboratory of Physics, Kochi University of Technology,
Tosa Yamada, Kochi 782-8502, Japan}

\date{\today}

\begin{abstract}
We design two simple quantum devices applicable as an adjustable quantum spectral filter and as a flux controller. Their function is based upon the threshold resonance in a F\"ul\"op-Tsutsui type star graph with an external potential added on one of the lines. Adjustment of the potential strength directly controls the quantum flow from the input to the output line. This is the first example to date in which the quantum flow control is achieved by addition of an external field not on the channel itself, but on other lines connected to the channel at a vertex.
\end{abstract}

\pacs{03.65.-w, 03.65.Nk, 73.63.Nm}
\maketitle


The use of quantum graphs as models of quantum devices is now being widely discussed 
\cite{AG05,EKST08}.
Quantum star graphs with $n\geq2$ lines, which are also ``elementary building blocks'' of any quantum graph, seem to serve particularly well for the purpose. They allow to design devices, that, although being technically simple, can have a wide scale of physical properties thanks to their large parameter spaces.
One of the first applications of quantum star graphs emerged in the spectral filtering. An $n=2$ star graph with the $\delta$-interaction in its center is already usable as a high-pass filter, and similarly, a graph with the $\delta'$-interaction works as a low-pass filter. Besides of these two simple designs, the existence of an $n=3$ branching filter, functionning as a high-pass/low-pass junction, has been proved \cite{CET09}.

In principle, such a system can be controlled by a variation of the vertex parameters. But this is hard to realize in practice since it requires real-time adjustments of a nanoscale object. It would be highly desirable if the control is achieved through an external field  applied onto one of the lines, preferably on lines other than those along which we want the quantum particles to propagate.

In this paper we show, for the first time, that a quantum filter controllable by an external potential can be indeed designed. Besides the filter, we construct one more similar device, namely a quantum ``sluice-gate'' which allows to increase and decrease the quantum flux from one line to another by adjusting the external potential applied to another line.
Our constructions are based on very simple star graphs with $n=3$ and $n=4$, respectively.
The presented result may serve also as a starting point in a search for other controllable quantum device models based on quantum graphs.

\medskip

When a quantum particle with mechanical energy $E$ living on a star graph comes in the vertex from the $j$-th line, it is scattered at the vertex into all the lines. The $i$-th component of the final-state wave function equals 
\begin{eqnarray}
\label{psi_ji}
\psi_{ij}(x)=\left\{\begin{array}{cl}
\frac{1}{\sqrt{k_j}}\e^{-{\rm i} k_j x}+\S_{jj}\frac{1}{\sqrt{k_j}}\e^{{\rm i} k_j x} & \quad\text{for } i=j\,, \\ \\
\S_{ij}\frac{1}{\sqrt{k_i}}\e^{{\rm i} k_i x} & \quad\text{for } i\neq j\,,
\end{array}\right. 
\end{eqnarray}
where $\S_{ij}$ are scattering amplitudes, $k_i$ are momenta on the corresponding lines, and coefficients $1/\sqrt{k_i}$ are involved for proper normalization. For any $i$, the momentum $k_i$ is equal to $k_i = \sqrt{E-U_i}$, where $U_i$ is the potential on the $i$-th line. The matrix $\S=\{\S_{ij}\}$ is the \emph{scattering matrix} of the graph. For a normalized wave function coming in from the $j$-th line, $\S_{ij}$ is interpreted as the complex amplitude of transmission into the $i$-th line (for $i\neq j$), whereas $\S_{jj}$ represents the complex amplitude of reflection. The matrix $\S$ depends, besides the internal properties of the vertex, on $E$ and $U_1,U_2,\ldots,U_n$.

To derive a formula for $\S$, let us define matrices $M=\{ \psi_{ij}(0)\}$ and $M'=\{ \psi'_{ij}(0)\}$. 
With regard to~\eqref{psi_ji}, it holds
\begin{eqnarray}
\label{M}
\!\!\!\!\!\!
M = K^{-1} + K^{-1} \S, \quad
M' = {\rm i} K (-K^{-1} + K^{-1} \S ),
\end{eqnarray}
where $K = \{ \sqrt{k_i}\delta_{ij} \}$. Any wave function $\Psi(x)=(\psi_1(x),\ldots,\psi_n(x))^T$ (the superscript $T$ denotes the transposition) on the graph obeys the \emph{boundary condition} determining the vertex, which is usually written in the form $A{\Psi}(0) + B\Psi'(0) = 0$ for certain fixed $A,B\in\C^{n,n}$, cf.~\cite{KS99}. In particular, b.~c. must be satisfied by the final-state wave function $\Psi_j(x)=(\psi_{1j}(x),\ldots,\psi_{nj}(x))^T$ determined in~\eqref{psi_ji} for all $j$, hence $AM + BM' = 0$, which together with~\eqref{M} leads to the sought expression for $\S$:
\begin{eqnarray}
\label{S}
\S = - (AK^{-1} + {\rm i}B K)^{-1}(AK^{-1} - {\rm i} B K)\,.
\end{eqnarray}
%
Squared moduli of the elements of $\S$ have the following interpretation: $|S_{ij}|^2$ for $j\ne i$ represents the probability of transmission 
from the $i$-th to the $j$-th line, $|S_{jj}|^2$ is the probability of reflection on the $j$-th line.

\smallskip

\begin{figure}[h]
  \centering
  \includegraphics[width=4.2cm]{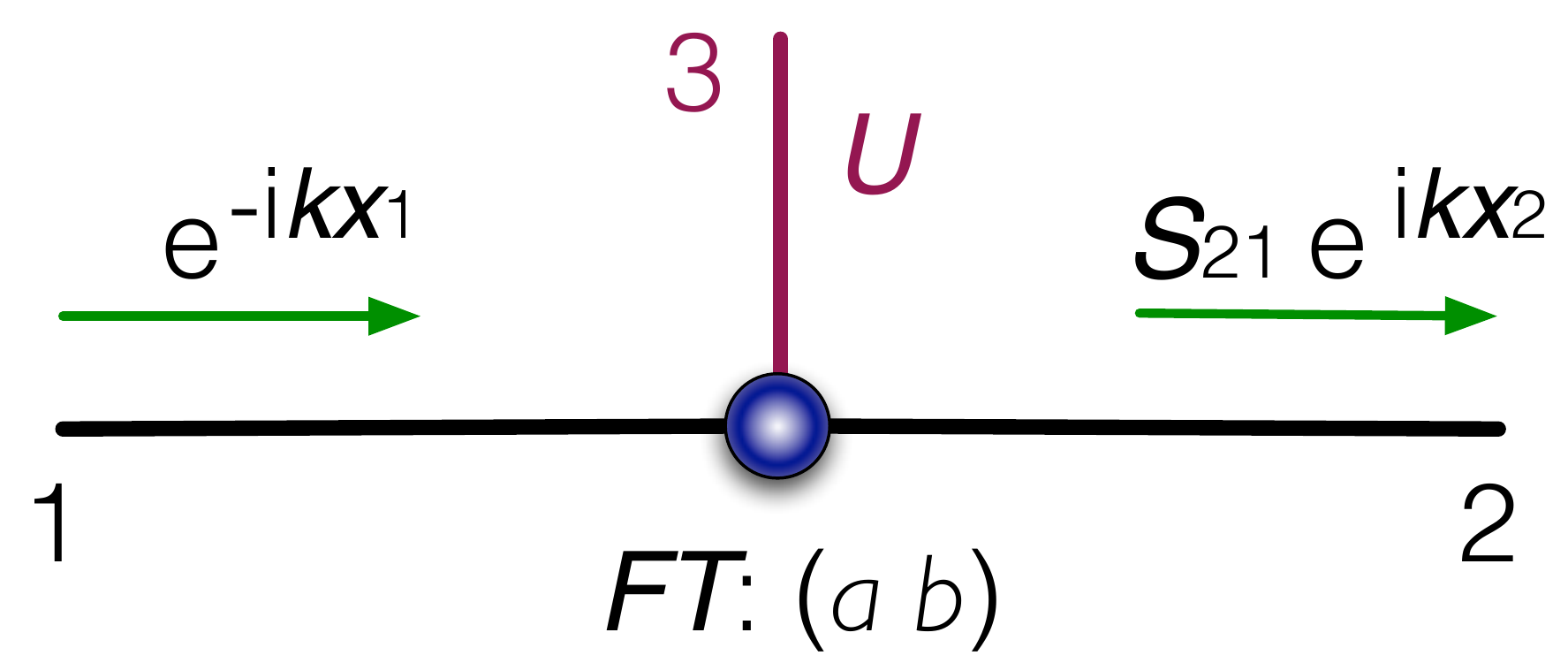}
  \caption{Schematic depiction of the $n=3$ star graph with an external potential $U$ on the line 3.}
  \label{fig:m1}
\end{figure}

Now consider an $n=3$ star graph with a F\"ul\"op-Tsutsui (also called ``scale invariant'') singular coupling (cf.~\cite{FT00, NS00, SS02}) in its vertex. For the sake of convenience, the coupling will be described by a boundary condition written in the so-called $ST$-form ($B\Psi'=-A\Psi$ with specially structured $A,B$, see~\cite{CET10} and~\cite{CT10}) with explicit notation
\begin{eqnarray}
\label{bc}
  \begin{pmatrix}
  1 & a & b \\ 0 & 0 & 0 \\ 0 & 0 & 0
  \end{pmatrix}
  \begin{pmatrix}
  \psi'_1(0) \\ \psi'_2(0) \\ \psi'_3(0)
  \end{pmatrix}
=
  \begin{pmatrix}
  0 & 0 & 0 \\ -a & 1 & 0 \\ -b & 0 & 1
  \end{pmatrix}
  \begin{pmatrix}
  \psi_1(0) \\ \psi_2(0) \\ \psi_3(0)
  \end{pmatrix}\,.
\end{eqnarray}
The graph is schematically illustrated in Fig.~\ref{fig:m1}.
The roles of individual lines are the following:
\begin{itemize}
\item Line 1 is \emph{input}. Particles of various energies are coming in the vertex along this line.
\item Line 2 is \emph{output}. Particles passed through the vertex are gathered on this line.
\item Line 3 is \emph{controling line}. We assume that this line is subjected to a constant (but adjustable) external potential $U$.
\end{itemize}
A quantum particle with energy $E=k^2$ coming in the vertex from the input line 1 is scattered at the vertex into all the lines. The scattering amplitudes can be calculated by substituting the matrices $A,B$ from the b.~c.~\eqref{bc} together with
\begin{eqnarray}
\label{}
k_1 = k_2 = k, \quad k_3 = \sqrt{k^2 - U}
\end{eqnarray}
into equation~\eqref{S}. We obtain:
\begin{eqnarray}
\label{S21}
\!\!\!\!\!\!\!\!\!\!\!\!
\S_{21}(k;U) = \frac{2a}{ 1+a^2+ b^2 \sqrt{1-\frac{U}{k^2} } }\,,
\end{eqnarray}
%
%
\begin{eqnarray}
\label{}
&&
\S_{11}(k;U) = \frac{1-a^2- b^2 \sqrt{1-\frac{U}{k^2} } }{ 1+a^2+ b^2 \sqrt{1-\frac{U}{k^2} } }\,,
\\
&&
\S_{31}(k;U) = \frac{2b  \left(1-\frac{U}{k^2} \right)^{\frac{1}{4}} \Theta(k-\sqrt{U}) }
{ 1+a^2+ b^2 \sqrt{1-\frac{U}{k^2} } }\,. \label{S31}
\end{eqnarray}
The Heaviside step function $\Theta(k-\sqrt{U})$ is added in equation~\eqref{S31} to make the expression valid for all energies $k^2$, including $k^2<U$. It represents asymptotically no transmission to the line 3 below the threshold momentum $k_\mathrm{th}=\sqrt{U}$.

\begin{figure}[h]
  \centering
  \includegraphics[width=6.5cm]{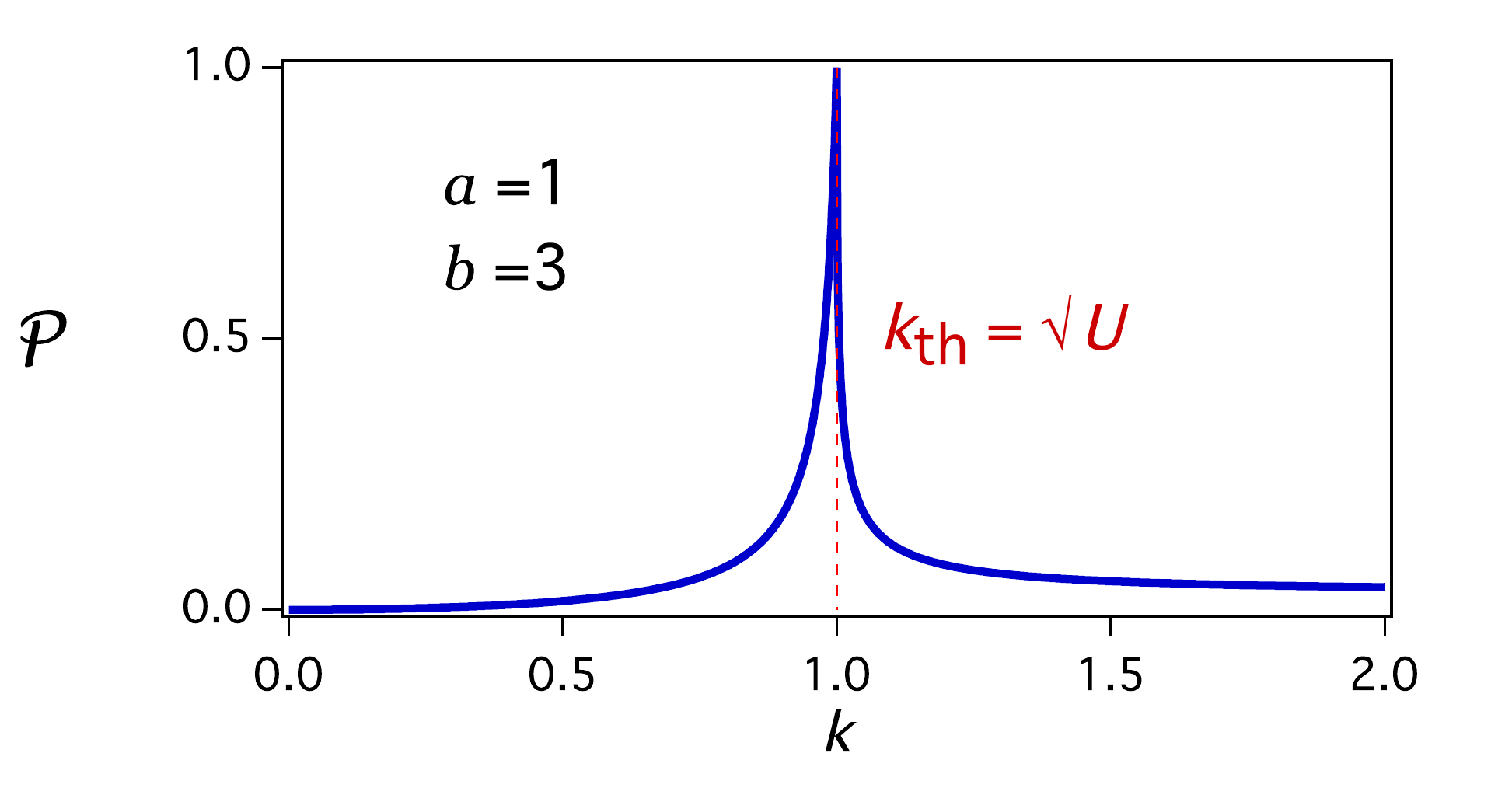}
  \includegraphics[width=6.5cm]{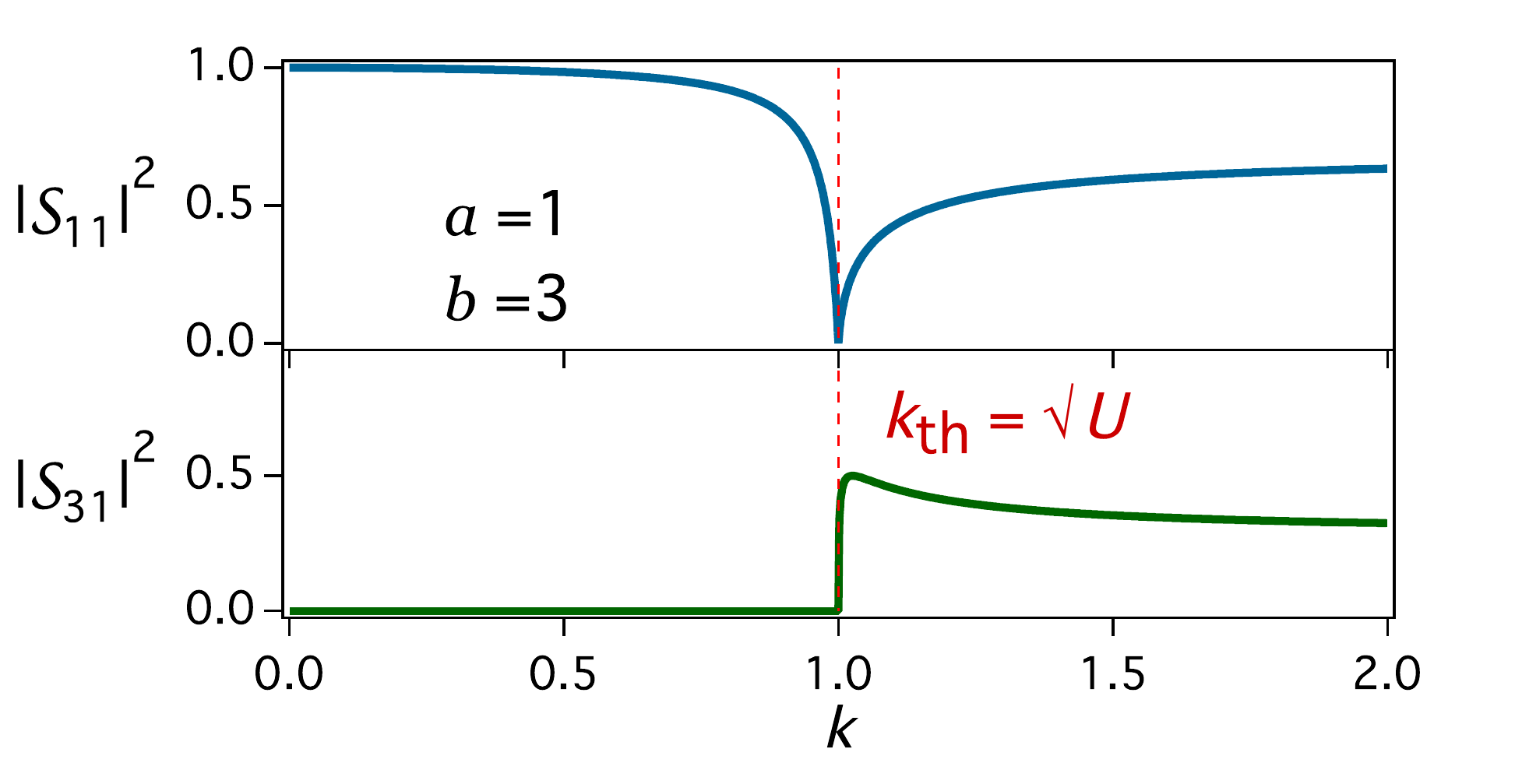}
  \caption{Scattering characteristics of the graph from Fig.~\ref{fig:m1} with parameters $a=1$, $b=3$. The transmission probability $\P(k;U)$ as a function of $k$ with the value of the potential set to $U=1$ is plotted in the top figure. The lower figure shows reflection probability $|\S_{11}(k; U)|^2$ and the probability of transmission to the controlling line $|\S_{31}(k; U)|^2$. }
  \label{fig:m2}
\end{figure}

We are interested in the probability of transmission into the output line $2$, which we denote by $\P(k;U)$, and particularly in its $k$-dependence. Since $\P(k;U)=|\S_{21}(k;U)|^2$,
we have from~\eqref{S21}
\begin{eqnarray}
\label{}
\!\!
\P(k;U)=\left\{\begin{array}{cl}
\frac{4a^2}{\left(1+a^2+b^2\sqrt{1-U/k^2}\right)^2} & \quad\text{for } k\geq\sqrt{U}, \\ \\
\frac{4a^2}{(1+a^2)^2+b^4(U/k^2-1)} & \quad\text{for } k\leq\sqrt{U}.
\end{array}\right.
\end{eqnarray}
We observe that for a given constant potential on the line 3, $\P(k;U)$ as a function of $k$ grows in the interval $(0,\sqrt{U})$, attains its maximum at $k=\sqrt{U}$, and decreases in the interval $(\sqrt{U},\infty)$. In particular, it holds:
\begin{eqnarray}
\label{}
&&
\lim_{k\to0}\P(k;U) = 0\,,
 \\
&&
\P(\sqrt{U};U) = \left(\frac{2a}{1 + a^2}\right)^2 \label{P(sqrt(U),U)}\,,
\\
&&
\lim_{k\to\infty}\P(k;U) = \left(\frac{2a}{1 + a^2 + b^2}\right)^2\,.
\end{eqnarray}
If the parameters $a,b$ satisfy $b \gg a\geq1$, the function $\P(k;U)$ has a sharp peak at
$k=\sqrt{U}$.
Equation~\eqref{P(sqrt(U),U)} implies that the peak is highest possible (attaining 1) for $a=1$. We conclude: If $b\gg a=1$, the system has high input$\to$output transmission probability for particles having momenta $k\approx\sqrt{U}$ and the transmission is perfect for $k=\sqrt{U}$, whereas there is just a very small transmission probability for other values of $k$. The situation is numerically illustrated in Fig.~\ref{fig:m2}. The quantum graph schematically depicted in Fig.~\ref{fig:m1} can be therefore used as an adjustable spectral filter, controllable by the potential put on the controlling line 3. The bandwidth $W$ of the filter, i.e., the width of the interval of energies $k^2$ for which $\P(k;U)>1/2$, depends on $U$ and $b$, and for $b\gg1$ it is approximately given as $W \approx 4.7 U/b^4$.
Let us remark that the resonance at the threshold momentum $k_\mathrm{th}=\sqrt{U}$ is related to the pole of the scattering matrix which is located on the positive real axis at
%
$k_{\mathrm{pol}}=\frac{b^2}{\sqrt{b^4-(1+a^2)^2}}\sqrt{U}$
%
on the unphysical Riemann surface which is connected to the physical Riemann surface at the cut that runs between $k=\pm\sqrt{U}$.  

\medskip

\begin{figure}[h]
  \centering
  \includegraphics[width=4.2cm]{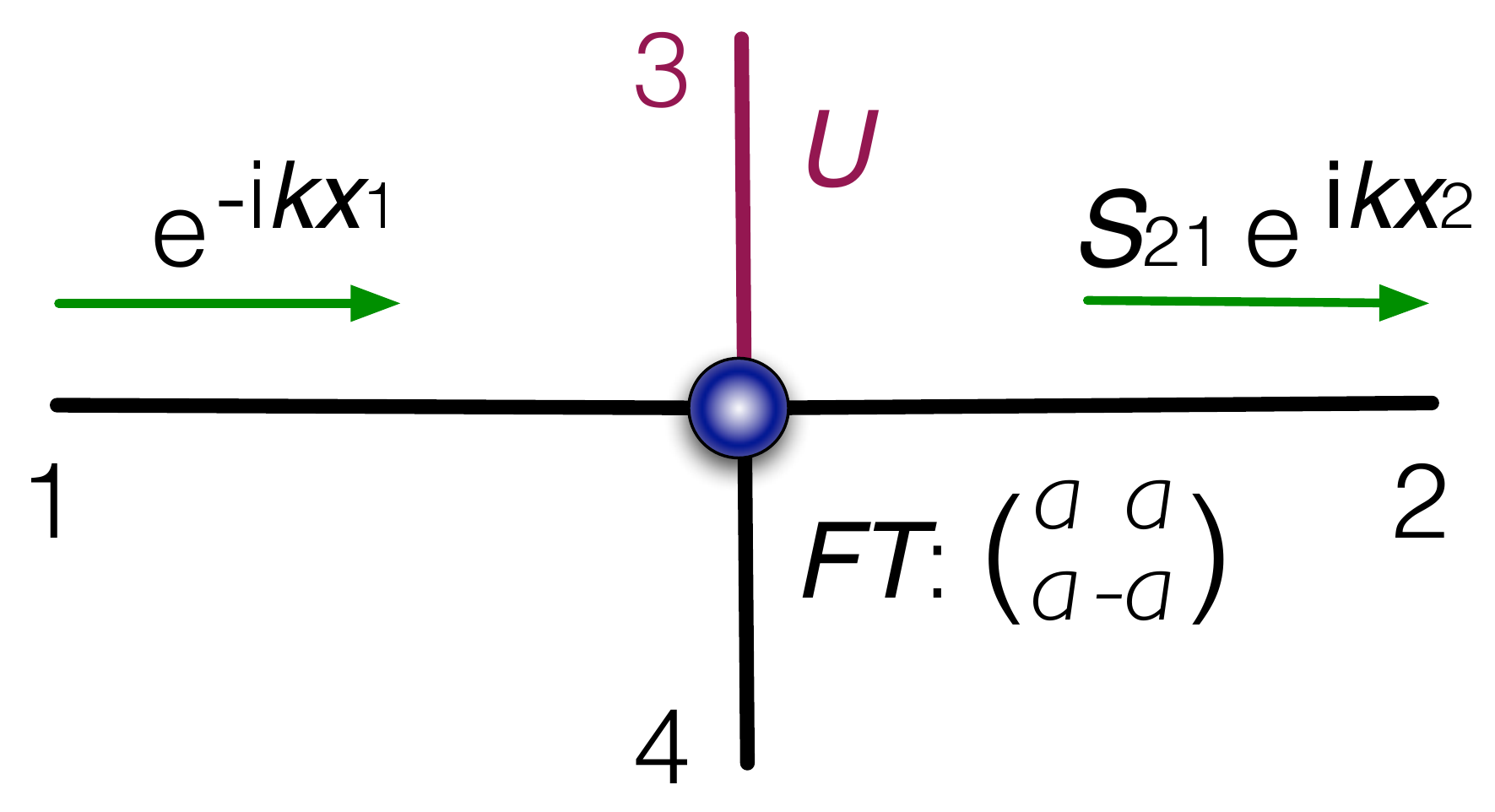}
  \caption{Schematic depiction of the $n=4$ star graph with an external potential $U$ on the line No. 3.}
  \label{fig:m3}
\end{figure}

In order to develop another quantum device, let us consider an $n=4$ star graph
as schematically illustrated in Fig.~\ref{fig:m3}. The meaning of the first three lines will be the same as in the previous model: 1 = input, 2 = output, 3 = controlling line subjected to a constant external potential $U$. The line No. 4 is a \emph{drain} and is included in the model for technical reasons: our considerations showed that the device we wish to construct is mathematically infeasible using a vertex of degree $n=3$.
The vertex coupling is again assumed to be of the F\"ul\"op-Tsutsui type, and its properties are determined by the boundary condition written for convenience in the $ST$-form
\begin{eqnarray}
\label{}
\!\!\!\!\!\!\!\!
  \begin{pmatrix}
  1 &  0 & a & a \\ 0 & 1 & a & -a \\ 0 & 0 & 0 & 0 \\ 0 & 0 & 0 & 0
  \end{pmatrix}\!\!
  \begin{pmatrix}
  \psi'_1(0) \! \\ \psi'_2(0) \! \\ \psi'_3(0) \! \\ \psi'_4(0) \! 
  \end{pmatrix}
\!=\!
  \begin{pmatrix}
  0 & 0 & 0 & 0 \\ 0 & 0 & 0 & 0 \\ -a & -a & 1 & 0 \\ -a & a & 0 & 1
  \end{pmatrix}\!\!
  \begin{pmatrix}
  \psi_1(0) \! \\ \psi_2(0) \! \\ \psi_3(0) \! \\ \psi_4(0) \!
  \end{pmatrix}
\end{eqnarray}
with $a\in\R$. The block $\begin{pmatrix} a & a \\ a & -a \end{pmatrix}$ is a special choice that ensued from our analysis; generally, the $ST$-form admits $\begin{pmatrix} a & b \\ c & d \end{pmatrix}$ for any $a,b,c,d\in\C$, cf.~\cite{CET10} or \cite{CT10}.

\begin{figure}[h]
  \centering
  \includegraphics[width=6.5cm]{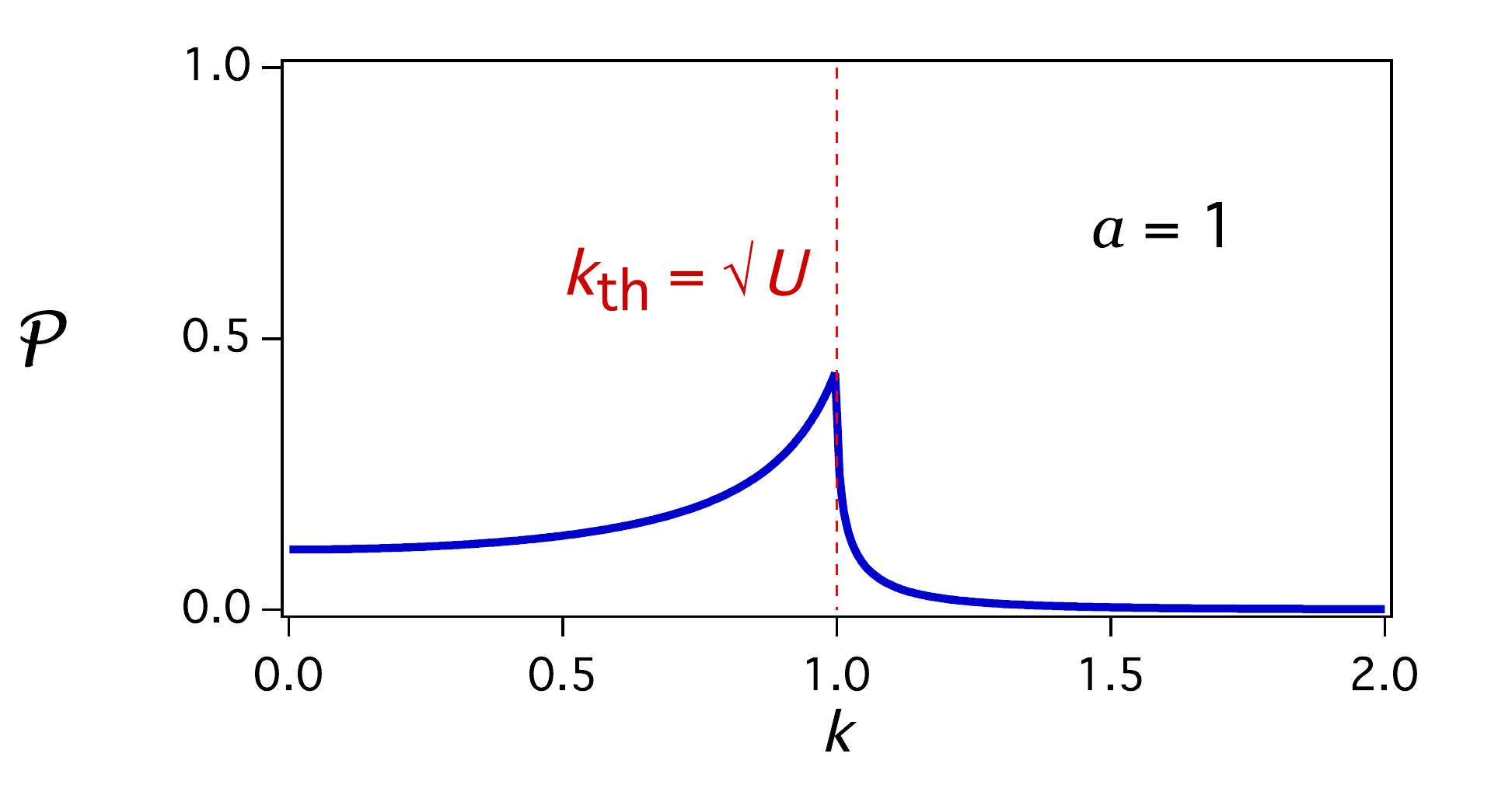}
  \includegraphics[width=6.5cm]{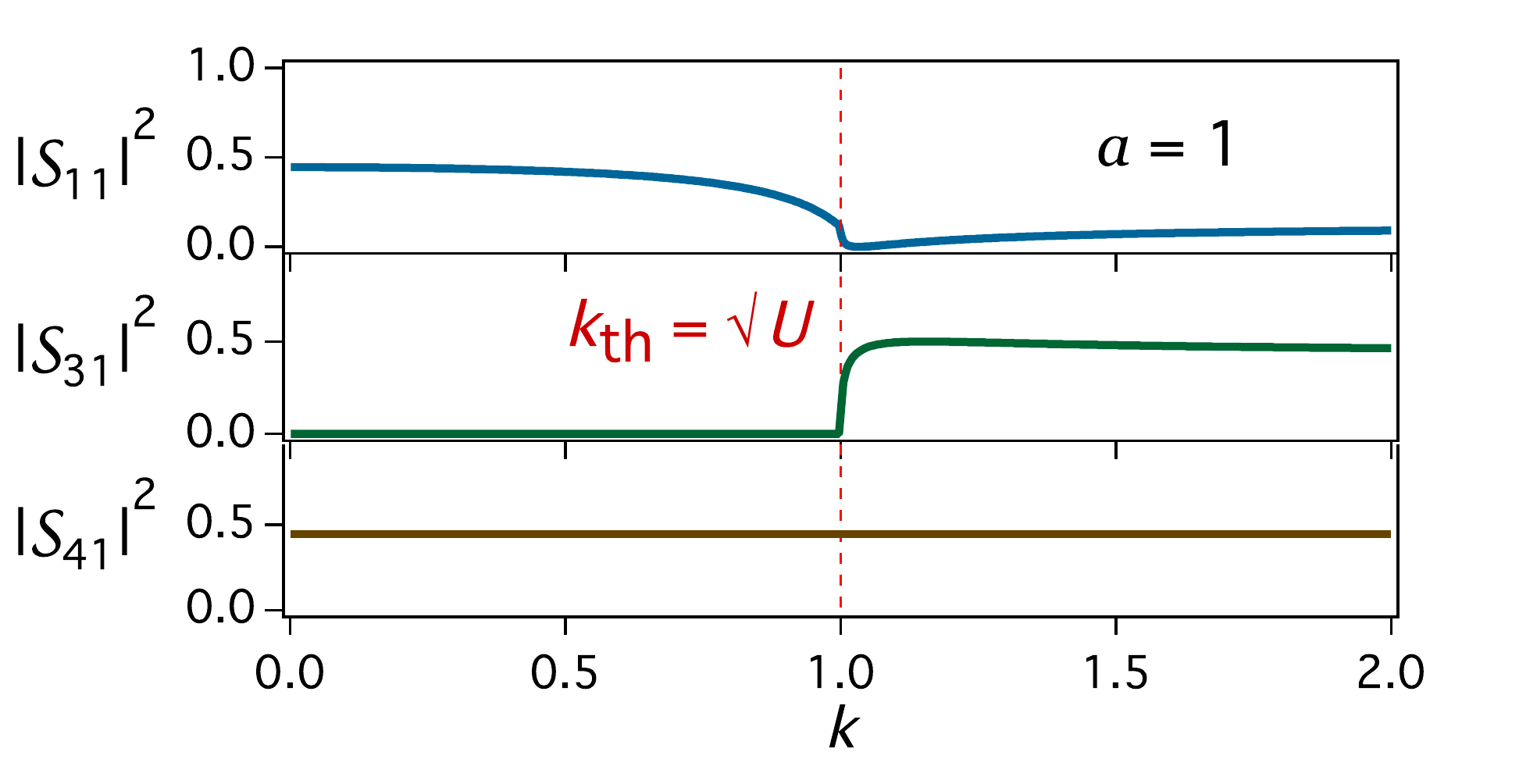}
  \caption{Scattering characteristics of the graph from Fig.~\ref{fig:m3} with parameter $a=1$. The transmission probability $\P(k;U)$ as a function of $k$ with the value of the potential set to $U=1$ is plotted in the top figure. The lower figure shows the reflection probability $|S_{11}(k; U)|^2$ and the probabilities of transmission to the controlling line $|S_{31}(k; U)|^2$ and to the drain line  $|S_{41}(k; U)|^2$. }
  \label{fig:4}
\end{figure}

\noindent
For a particle with energy $E=k^2$ coming in the vertex from the input line $1$, we have
\begin{eqnarray}
\label{}
k_1 = k_2 = k, \quad k_3 = \sqrt{k^2 - U}, \quad k_4 = k,
\end{eqnarray}
and the scattering amplitudes can be calculated as
\begin{eqnarray}
\label{}
\!\!\!\!\!\!\!\!\!\!\!\!
S_{21}(k;U) = 
\frac{ 2a^2\left( 1-\sqrt{1-\frac{U}{k^2} } \right) }{  (1 + 2a^2)  + 2a^2(1+2a^2) \sqrt{1-\frac{U}{k^2}}  }
\end{eqnarray}
and
\begin{eqnarray}
\label{}
&&\!\!\!\!\!\!\!\!\!\!\!\!
S_{11}(k;U) = \frac{ 1-4a^4  \sqrt{1-\frac{U}{k^2} } }
{  (1 + 2a^2)  + 2a^2(1+2a^2) \sqrt{1-\frac{U}{k^2}}  }\,,
\\
&&\!\!\!\!\!\!\!\!\!\!\!\!
S_{31}(k;U) = \frac{ 2a(1+2a^2)  \left(1-\frac{U}{k^2} \right)^{\frac{1}{4}}  \Theta(k-\sqrt{U})  }
{  (1 + 2a^2)  + 2a^2(1+2a^2) \sqrt{1-\frac{U}{k^2}}  }\,,
\\
&&\!\!\!\!\!\!\!\!\!\!\!\!
S_{41}(k;U) = \frac{ 2a + 4a^3\sqrt{1-\frac{U}{k^2}} }
{  (1 + 2a^2)  + 2a^2(1+2a^2) \sqrt{1-\frac{U}{k^2}}  }\,.
\end{eqnarray}
We again denote the transmission probability input$\to$output by $\P(k;U)=|\S_{21}(k;U)|^2$. It holds
\begin{eqnarray}
\label{}
\!\!\!\!\!\!\!\!\!\!
\P(k;U)=\left\{\begin{array}{ll}
\frac{4a^4U/k^2}{(1+2a^2)^2\left(1-4a^4+4a^4U/k^2\right)} & \text{ for } k\leq\sqrt{U}, \\  \\
\frac{4a^4\left(1-\sqrt{1-U/k^2}\right)^2}{(1 + 2a^2)^2\left(1+2a^2\sqrt{1-U/k^2}\right)^2} & \text{ for } k\geq\sqrt{U},
\end{array}\right.
\end{eqnarray}
hence
\begin{eqnarray}
\label{}
&&
\lim_{k\to0}\P(k;U) = \frac{1}{(1+2a^2)^2}\,,
 \\
&&
\P(\sqrt{U};U) = \frac{4a^4}{(1+ 2a^2)^2} \label{P4(sqrt(U),U)}\,,
\\
&&
\lim_{k\to\infty}\P(k;U) = 0\,.
\end{eqnarray}
If $U$ is fixed, $\P(k;U)$ as a function of $k$ quickly falls off to zero at $k>\sqrt{U}$. A typical behaviour is illustrated in a numerical example in Fig.~\ref{fig:4}. 
The peak at the threshold momentum $k_\mathrm{th}=\sqrt{U}$, appearing for $a>1/\sqrt{2}$, is again related to the pole in the unphysical Riemann plane at 
%
$k_{\mathrm{pol}}=\frac{2a^2}{\sqrt{(4a^4-1)}}\sqrt{U}$.
%

\begin{figure}[h]
  \centering
  \includegraphics[width=6.5cm]{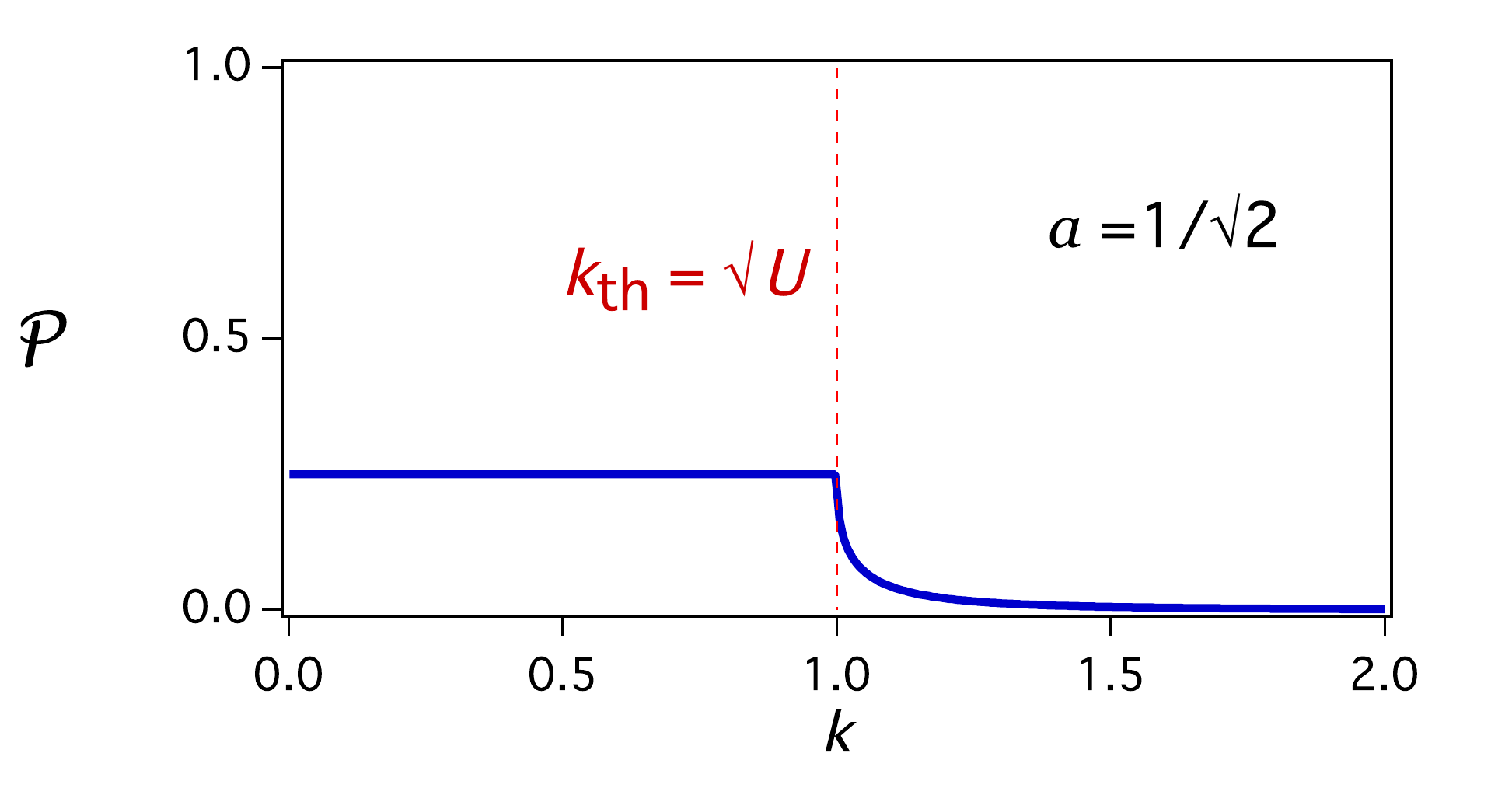}
  \includegraphics[width=6.5cm]{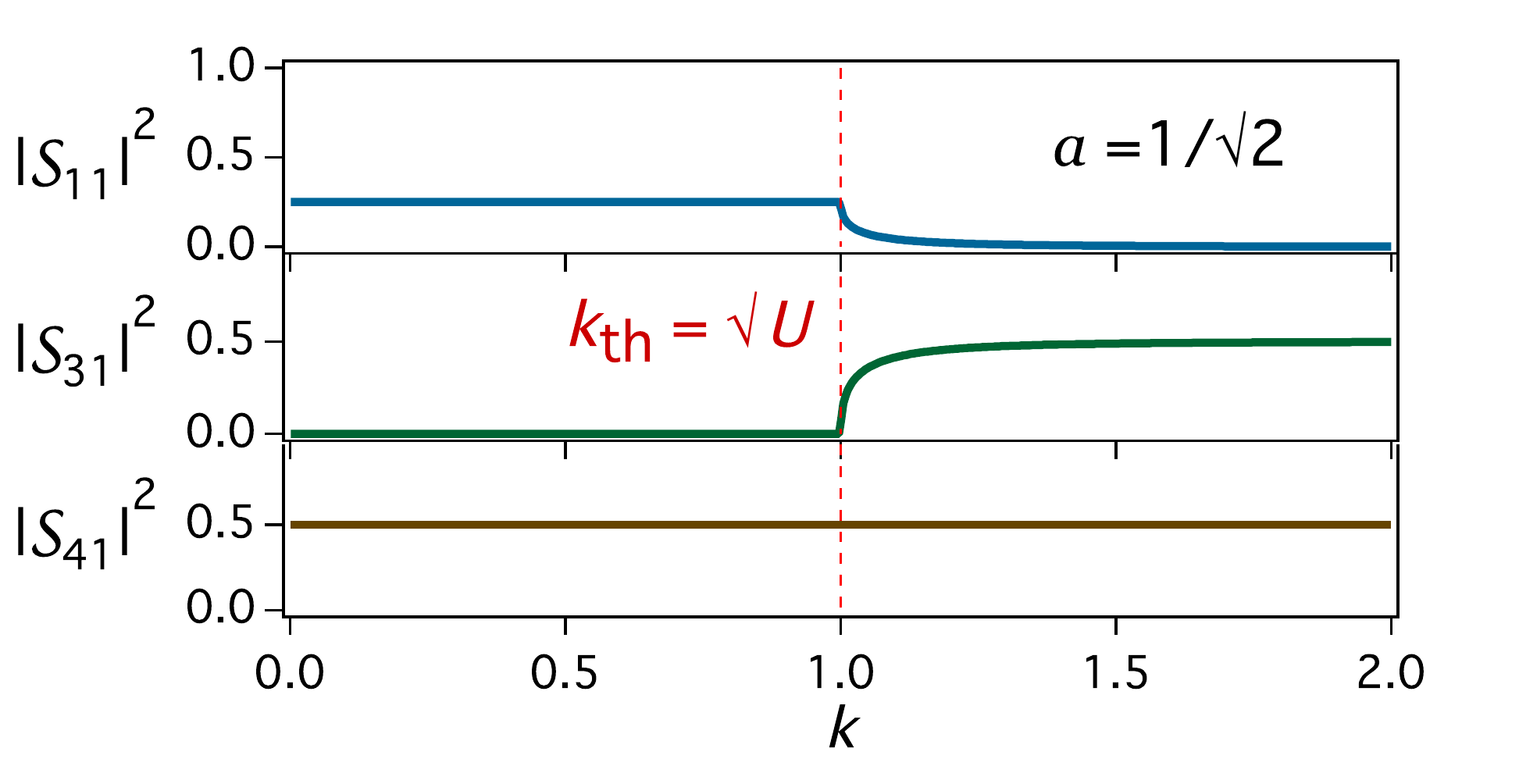}
  \caption{Characteristics of the flat spectral filter obtained from the graph on Fig.~\ref{fig:m3} for $a=1/\sqrt{2}$. The transmission probability $\P(k;U)$ as a function of $k$ with the value of the potential set to $U=1$ is plotted in the top figure. The lower figure shows the reflection probability $|S_{11}(k; U)|^2$ and the probabilities of transmission to the controlling line $|S_{31}(k; U)|^2$ and to the drain line  $|S_{41}(k; U)|^2$.}
  \label{fig:5}
\end{figure}

There is a value of the parameter $a$ that deserves a particular attention, namely $a=1/\sqrt{2}$. For this choice of $a$ the peak disappears and the function $\P(k;U)$ becomes constant in the whole interval $(0,\sqrt{U})$:
\begin{eqnarray}
\!\!\!\!\!\!\!\!\!
\P(k;U)=\left\{\begin{array}{cl}
\frac{1}{4} & \quad\text{for } k\leq\sqrt{U}, \\ \\
\frac{1}{4}\cdot\left(\frac{1-\sqrt{1-U/k^2}}{1+\sqrt{1-U/k^2}}\right)^2 & \quad\text{for } k>\sqrt{U},
\end{array}\right.
\end{eqnarray}
see Fig.~\ref{fig:5}. The device then behaves as a spectral filter with a flat passband that transmits (with the probability of $1/4$) quantum particles with momenta $k \in [0,\sqrt{U}]$ to the output, whereas particles with higher momenta are diverted to other lines, mainly to 3 and 4. The process is directly controlled by the external potential $U$.
Since increasing $U$ opens the channel 1$\to$2 for more particles, the device can be regarded as a quantum sluice-gate, applicable as a quantum flux controller.
When there are many particles described by the momentum distribution $\rho(k)$ on the line 1, the flux $J$ to the line 2 is given by
%
$
J(U) = \int dk \rho(k) k \P(k;U)
$.
%
Assuming the Fermi distribution with Fermi momentum $k_F$ larger than our range of operation of $\sqrt{U}$, we can set $\rho(k)=\rho=\mathrm{const}$. With the approximation $\P(k;U) \approx \frac{1}{4} \Theta(\sqrt{U}-k)$, we obtain $J(U) = \frac{1}{8}\rho U$, which indicates the linear flux control.

The sluice-gate built from an $n=4$ star graph has one more operation mode. If the line No. 4 (the drain) is subjected to another external field $V$, $0<V<U$, the channel 1$\to$2 opens for particles with $k \in [\sqrt{V},\sqrt{U}]$ and mostly closes for particles with $k$ outside this interval. The gate then works as a fully tunable band spectral filter. However, in contrast to the standard $V=0$ operation mode, the filter with $V>0$ does not have a flat passband.

\begin{figure}[t]
  \centering
  \includegraphics[width=4.1cm]{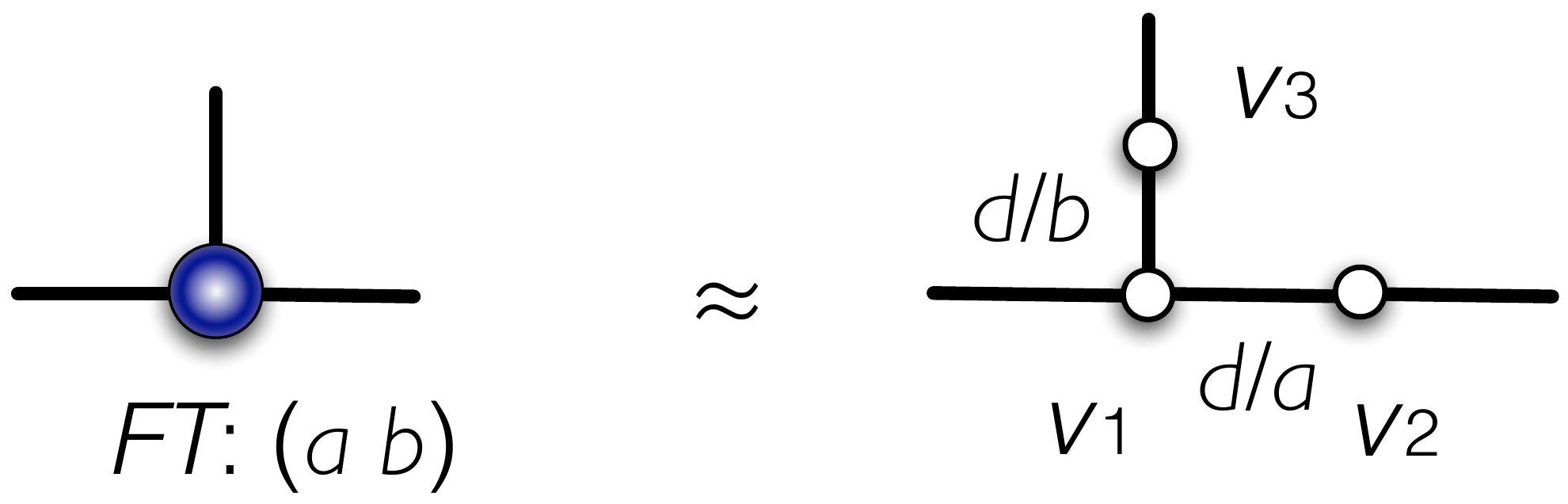}
\\ \smallskip
  \includegraphics[width=4.1cm]{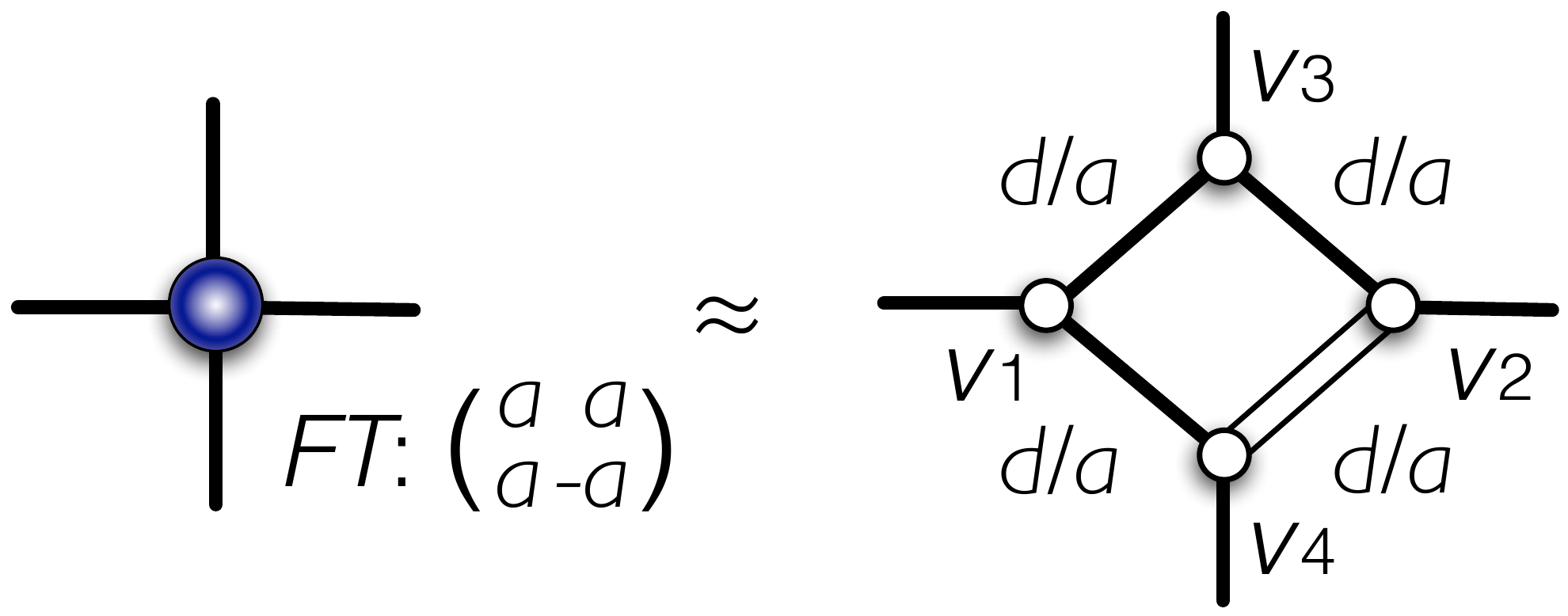}
  \caption{Finite constructions of the F\"ul\"op-Tsutsui couplings used. The design, based on~\cite{CT10}, utilizes the $\delta$-couplings connected by short lines. The small size limit $d\to 0$ with the $\delta$-coupling strengths scaled with $d$ effectively produces the required F-T vertex coupling. For the $n=3$ case (top), the $\delta$-coupling strengths are given by $v_1=[a(a-1)+b(b-1)]/d$, $v_2=(1-a)/d$ and $v_3=(1-b)/d$.
  For the $n=4$ case (bottom), the strengths are $v_1=v_2=2a(a-1)/d$,  $v_3= v_4 =(1-2a)/d$. The double line represents a line with a ``magnetic'' vector potential, which can be alternatively replaced by a line carrying the $\delta$-coupling of strength $v_5=-8a/d$ in its center, together with changing $v_2$ and $v_4$ to $v_2=2a(a-2)/d$, $v_4 =(1-4a)/d$. }
  \label{fig:6}
\end{figure}

%
\smallskip
It should be emphasized that the studied controllable filter devices using the threshold resonance became possible only with ``exotic'' F\"ul\"op-Tsutsui-type couplings in the vertices. Standard vertex couplings (the free and the $\delta$-coupling) would not work this way. It is therefore essential, for the proposed designs to be viable, that the required F\"ul\"op-Tsutsui vertices can be created using standard couplings, which themselves have a simple physical interpretation \cite{Ex96b}. This problem has been addressed in \cite{CET10} and \cite{CT10}, where it was proved that any F\"ul\"op-Tsutsui coupling given by b.~c. with real matrices $A,B$ can be approximately constructed by assembling a few $\delta$-couplings. The solution for our case is shown in Fig.~\ref{fig:6}.
%

\medskip
\smallskip
We thank Prof. E. F. Redish for helpful suggestions.  This research was supported  by the Japan Ministry of Education, Culture, Sports, Science and Technology under the Grant number 21540402.

\end{document}